\documentclass[runningheads,fleqn]{svmult}
\usepackage{makeidx}   % allows index generation \usepackage{graphicx}
\usepackage{subeqnar}  % subnumbers individual equations
\usepackage{multicol}  % used for the two-column index
\usepackage{taphys}    % flushleft layout of captions etc.
\usepackage{braket} 
\usepackage{graphicx}
\makeindex             % used for the subject index
 \newcommand{\up}{\uparrow}
\newcommand{\down}{\downarrow}

\begin{document}
\title*{Decoherence of Flux Qubits Coupled to Electronic Circuits}
\toctitle{Decoherence of Flux Qubits Coupled to Electronic Circuits}
% allows explicit linebreak for the table of content
%
%
\titlerunning{Decoherence of Flux Qubits Coupled to Electronic Circuits}
% allows abbreviation of title, if the full title is too long
% to fit in the running head
%
\author{F.K. Wilhelm\inst{1} \and M.J. Storcz\inst{1} \and C.H. van
der Wal\inst{2} \and C.J.P.M. Harmans\inst{3} \and J.E. Mooij\inst{3}}
\authorrunning{F.K. Wilhelm et al.}
% if there are more than two authors,
% please abbreviate author list for running head
%
%
\institute{Sektion Physik and CeNS, Ludwig-Maximilians-Universit\"at,
80333 M\"unchen, Germany \and Dpt.\ of Physics, Harvard University,
Cambridge, MA 02138, USA \and Dpt.\ of Nanoscience, Delft University
of Technology, 2600  GA Delft, Netherlands}

\maketitle              % typesets the title of the contribution

\begin{abstract}
On the way to solid-state quantum computing, overcoming
decoherence is the central issue. In this contribution, we discuss
the modeling of decoherence of a superonducting flux qubit coupled to 
dissipative electronic circuitry. We discuss its impact on single qubit
decoherence rates and on the performance of two-qubit gates.  These
results can be used for designing decoherence-optimal setups.
\end{abstract}

Quantum computation is one of the central interdisciplinary research
themes in present-day physics \cite{Bouwmeester}. It promises a
detailed  understanding of the often counterintuitive predictions of
basic quantum  mechanics as well as a qualitative speedup of certain
hard computational problems. A generic, although not necessarily
exclusive, set of  criteria for building quantum computers  has been
put forward by DiVincenzo \cite{DiVin}.  The experimental realization
of quantum bits has been pioneered in atomic physics,
optics and NMR. There,   the approach
is taken to use microscopic degrees of freedom which are well isolated
and can be kept quantum coherent over long times.  Efficient
controls are attached to these degrees of freedom. Even though these
approaches are immensely succesful demonstrating elementary operations, it
is not evident how they can be scaled up to macroscopic computers.

Solid-state systems on the other hand have proven to be scalable 
in present-day classical computers. Several proposals for
solid-state based {\em quantum} computers have
been put forward, many of them in the context of superconductors
\cite{Makhlin}.
As solid-state systems
contain a macroscopic number of degrees of freedom, they are very
sensitive to decoherence. Mastering and optimizing this decoherence is
a formidable task and requires deep understanding of the physical
system under investigation. Recent experimental success
\cite{Irinel,Yasu} suggests that this task can in principle be
performed.

In this contribution, we 
are going to study
decoherence of superconducting qubits coupled to an electromagnetic
environment which produces Johnson-Nyquist noise. We show, how the 
decoherence properties can be engineered
by carefully designing the environmental impedance. We will discuss how
the decoherence affects the performance of a CNOT operation.

\section{Superconducting Flux Qubits}

Superconducting qubits \cite{Makhlin,Irinel,Yasu,Hans} are very well
suited for the task of solid-state quantum computation, because two of the most obvious decoherence
sources in  solid-state systems are supressed: Quasiparticle
excitations experience an energy gap and phonons are frozen out at low
temperatures \cite{Lin}. The  computational Hilbert space is engineered using
Josephson tunnel junctions that are characterized by two competing
energy scales: The Josephson coupling of a junction 
with critical current $I_{\rm c}$, $E_{\rm J}=I_{\rm c}\Phi_0/2\pi$, 
and the charging energy $E_{\rm ch}=2e^2/C_{\rm J}$ of a
single Cooper pair on the geometric capacitance $C_{\rm J}$ of the junction. 
Here $\Phi_0=h/2e$ is the superconducting flux quantum. There
is a variety of qubit proposals classified by the ratio of this
energies. Whereas another contribution in this volume
\cite{MakhlinHere} focuses on the case of charge qubits, 
$E_{\rm ch}>E_{\rm J}$, this contribution is
motivated by flux qubit physics, $E_{\rm J}>E_{\rm ch}$.
However, most of the
discussion has its counterpart in other superconducting setups as
well.
\begin{figure}[htb]
%h=here, t=top, b=bottom, p=separate figure page
%\begin{center}\leavevmode
\includegraphics[width=0.8\columnwidth]{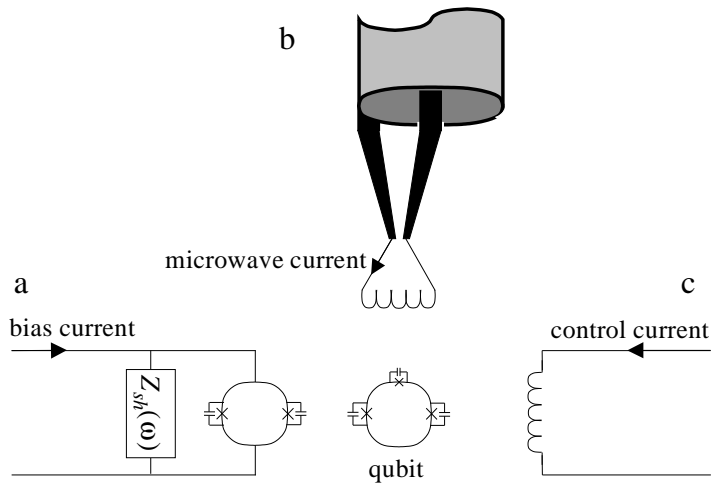}
\caption{Experimental setup for measurements on a 
flux qubit. The qubit (center) is a superconducting
loop that contains three Josephson junctions. It is inductively
coupled to a DC-SQUID (a), and superconducting control lines for
applying magnetic fields at microwave frequencies (b) and static
magnetic fields (c). The DC-SQUID is realized with an on-chip
shunt circuit with impedance $Z(\omega)$. The circuits a)-c) are
connected to filtering and electronics (not drawn)}
\label{figexp}
%\end{center}
\end{figure}
Specifically, we discuss a three junction qubit \cite{Hans,Caspar}, a
micrometer-sized
low-inductance superconducting loop containing three Josephson
tunnel junctions (Fig.~\ref{figexp}). By applying an external flux
$\Phi _{q}$ a persistent supercurrent can be induced in the loop. For
values where $\Phi _{q}$ is close to a half-integer number of
flux quanta, two states with persistent
currents of opposite sign are nearly degenerate but separated by an
energy barrier. We will assume here that the system is operated near
$\Phi _{q}=\frac12 \Phi _{0}$. The persistent currents in the
classically stable states
have here a magnitude $I_{\rm p} $. Tunneling through the barrier causes a
coupling between the two states, and at low  energies the loop
can be described by a Hamiltonian of a two state system
\cite{Hans,Caspar},
\begin{equation}
\hat{H}_{q}=\frac{\varepsilon }{2}\hat{\sigma}_{z}+\frac{\Delta
}{2}\hat{\sigma}_{x},
\label{2lsHam}
\end{equation}
where $\hat{\sigma}_{z}$ and $\hat{\sigma}_{x}$ are Pauli matrices. 
The two eigenvectors of $\hat{\sigma}_{z}$ correspond to
states that have a left or a right circulating current and will be
denoted as $|L\rangle $ and $|R\rangle $. The energy bias $\varepsilon
= 2I_{p}(\Phi _{q}-\frac12 \Phi _{0})$ is controlled by the externally
applied field $\Phi _{q}$. We follow \cite{grifoni99} and define
$\Delta $ as the tunnel splitting at $\Phi _{q}= \frac12 \Phi _{0}$,
such that $\Delta =2W$ with $W$ the tunnel coupling between the
persistent-current states. This system has two energy eigen values
$\pm \frac{1}{2}\sqrt{\Delta ^{2}+\varepsilon ^{2}}$, such that the
level separation $\nu $ gives $\nu =\sqrt{\Delta ^{2}+\varepsilon ^{2}}$.
In general $\Delta$ is a function of $\varepsilon$.  However, it
varies on the scale of the single junction plasma frequency, which is
much above the typical energy range at which the qubit is operated,
such  that we can assume $\Delta$ to be constant for the purpose of
this paper.

In the experiments $\Phi _{q}$ can be controlled by applying a
magnetic field with a superconducting coil at a distance
from the qubit and for local control one can apply currents to
superconducting lines, fabricated on-chip in the 
vicinity of the qubit. The qubit's quantum dynamics will be controlled
with resonant microwave pulses (i.~e.~by Rabi oscillations). In recent
experiments the qubits were operated at $\varepsilon\approx 5\Delta$ or
$\varepsilon\approx0$ \cite{Irinel,Caspar}. The numerical values
given in this paper will concentrate on the former case. At this point, there
is a good  trade-off between a system with significant
tunneling, and  a system with $\hat{\sigma}_{z}$-like eigenstates that can
be used for qubit-qubit couplings and measuring qubit states
\cite{Hans}.  The qubit has a magnetic dipole moment as a result of the
clockwise or counter-clockwise persistent current The corresponding
flux in the loop is much smaller than the  applied flux $\Phi_{q}$,
but large enough to be detected with a SQUID.  This will be used for
measuring the qubit states. For our two-level system Eq.\ (\ref{2lsHam}),
this means that  both manipulation and readout couple to
$\hat{\sigma}_{z}$. Consequently, the Nyquist noise produced by the necessary
external circuitry will couple in as flux noise and hence  couple to
$\hat{\sigma}_{z}$, giving $\epsilon$ a small, stochastically
time-dependent part $\delta\epsilon (t)$.  

Operation at $\varepsilon\approx0$ has the advantage that the flux
noise leads to less variation of $\nu$. 
In the first experiments \cite{Irinel} this has turned out to be crucial
for observing time-resolved quantum dynamics. Here, the qubit states can
be measured by incorporating the qubit inside the DC-SQUID loop. While
not working that out in detail, the methods that we present can also be 
applied for the analysis of this approach. This also applies to
the analysis of the impact of 
electric dipole moments, represented by $\hat{\sigma}_{x}$. With
$E_{\rm ch}\ll E_J$, these
couple much less to the circuitry and will hence not be discussed
here.

As the internal baths are well suppressed, the coupling to
the electromagnetic environment (circuitry, radiation noise) becomes a
dominant source of decoherence. This is a subtle issue: It is
not possible to couple the circuitry arbitrarily weakly or seal the
experimental setup, because it has to remain possible to
control the system. One rather has to engineer the electromagnetic
environment to combine good control with low unwanted
back-action.

Any linear electromagnetic environment can be described by an
effective impedance $Z_{\rm eff}$. If the circuit contains Josephson
junctions below their critical current, they
can be included through their kinetic inductance 
$L_{\rm kin}=\Phi_0/(2\pi I_{\rm c}\cos\bar{\phi})$,  where
 $\bar{\phi}$ is the average phase
drop across the junction. The circuitry disturbs the qubit through its
Johnson-Nyquist noise, which has Gaussian statistics and can thus be
described by an effective Spin-Boson model \cite{LeggettReview}.  In
this model, the properties of the oscillator bath which forms the
environment are characterized through a spectral function $J(\omega)$,
which can be derived from the external impedance. Note, that other
nonlinear elements such as tunnel junctions which can produce
non-Gaussian shot noise are generically {\em not} covered by
oscillator bath models.

As explained above, the flux noise from an external circuit leads to
$\epsilon=\epsilon_0+\delta\epsilon(t)$ in Eq. (\ref{2lsHam}).
We parametrize the noise
$\delta\epsilon(t)$ by its power spectrum
\begin{equation}
\left\langle\left\lbrace
\delta\epsilon(t),\delta\epsilon(0)\right\rbrace
\right\rangle_\omega=\hbar^2J(\omega)\coth(\hbar\omega/2k_BT).
\end{equation}
Thus, from the noise properties calculated by other means one can find
$J(\omega)$ as was explained in Detail in \cite{EPJB}. 
In this contribution, we would like to outline an alternative approach
pioneered by Leggett \cite{Leggett}, where $J(\omega)$ is derived
from the classical friction induced by the environment. In reality,
the combined system of SQUID and qubit will experience fluctuations arising
from additional circuit elements at different temperatures, which can
be treated in a rather straightforward manner.

\section{Decoherence from the Electromagnetic Environment}

\subsection{Characterizing the Environment from Classical Friction}

We study a DC-SQUID in an electrical circuit as shown in Fig.\
\ref{figexp}. It contains two Josephson junctions with phase drops
denoted by $\gamma_{1/2}$.
We start by looking at the average phase
$\gamma_{\rm ex}=(\gamma_1+\gamma_2)/2$  across the read-out
SQUID. Analyzing the circuit with Kirchhoff rules,  we find the
equation of motion
\begin{equation}
2C_J \frac{\Phi_0}{2\pi}\ddot{\gamma}_{\rm ex}=-2I_{c,0} \cos
(\gamma_i)\sin\gamma_{\rm ex} +I_{\rm bias}-\frac{\Phi_0}{2\pi}\int
dt^\prime \dot{\gamma}_{\rm ex}(t^\prime) Y(t-t^\prime).
\label{eq:tiltwashboard}
\end{equation}
Here, $\gamma_{\rm in}=(\gamma_1-\gamma_2)/2$ is the dynamical variable
describing the circulating current in the loop which is controlled by 
the flux,
$I_{\rm bias}$ is the bias current imposed by the source,
$Y(\omega)=Z^{-1}(\omega)$ is the admittance in  parallel to the whole
SQUID and $Y(\tau)$ its Fourier transform. The SQUID is described by
the junction critical currents $I_{\rm c,0}$ which are assumed to be
equal, and their capacitances $C_{\rm J}$. We now proceed by finding 
a static solution which sets the operation point $\gamma_{\rm in/ex,0}$ 
 and small fluctuations around them, $\delta\gamma_{\rm in/ex}$.
The static solution
reads $I_{\rm bias}=I_{\rm c,eff}\sin\gamma_{\rm ex,0}$
where $I_{\rm c,eff}=2I_{c,0}\cos\gamma_{in ,0}$ is the effective
critical current of the SQUID. Linearizing Eq. \ref{eq:tiltwashboard}
around this solution and Fourier-transforming,  we find that
\begin{equation}
\delta\gamma_{\rm ex}(\omega)=\frac{2\pi I_{\rm b}\tan\gamma_{\rm in,0} Z_{\rm
eff}(\omega)}{i\omega\Phi_0} \delta\gamma_i(\omega)
\label{eq:linext}
\end{equation}
where $Z_{\rm eff}(\omega)=\left(Z(\omega)^{-1}+2i\omega C_{\rm
J}+(i\omega L_{\rm kin})^{-1}\right)^{-1}$ is the effective impedance
of the  parallel circuit consisting of the $Z(\omega)$, the kinetic
inductance of the SQUID and the capacitance of its
junctions. Neglecting self-inductance of the SQUID and the
(high-frequency) internal plasma mode, we can  straightforwardly
substitute $\gamma_{\rm in}=\pi \Phi/\Phi_0$ and  split it into $\gamma_{\rm
in,0}=\pi\Phi_{\rm x,S}/\Phi_0$ set by the externally  applied flux
$\Phi_{\rm x,S}$ through the SQUID loop  and $\delta\gamma_{\rm i}=\pi
M_{\rm SQ}I_{\rm Q}/\Phi_0$ where $M_{\rm SQ}$ is the mutual inductance between
qubit and the SQUID and $I_{\rm Q}(\vec{\varphi})$ 
is  the circulating current in the qubit
as a function of the junction phases, which assumes values $\pm I_{\rm p}$ 
in the classically stable states.

In order to analyze the backaction of the SQUID onto 
the qubit in the two-state approximation, 
Eq.\ (\ref{2lsHam}), we have to get back to its full, continuous description,
starting from the classical dynamcis. These are equivalent to 
a particle, whose coordinates are the two independent
junction phases in the three-junction loop, in a two-dimensional
potential 
\begin{equation}
\vec{C}(\Phi_0/2\pi)^2\ddot{\vec{\varphi}}=-\nabla U(\vec{\varphi},
\Phi_{x,q}+I_{\rm S}M_{\rm SQ}).
\end{equation}
The details of this equation are explained in \cite{Hans}. 
${\vec C}$ is the capacitance matrix describing the charging
of the Josephson junctions in the loop, $U(\vec{\varphi})$ contains
the Josephson energies of the junctions as a function of the junction
phases and
$I_{\rm S}$.
is the ciculating current in the SQUID loop. 
The applied flux through the qubit $\Phi_{\rm q}$ is split into the flux
from the external coil $\Phi_{x,q}$ and the contribution form the 
SQUID. 
Using the above relations we find
\begin{equation}
I_{\rm S}M_{\rm SQ}=\delta\Phi_{\rm cl}- 2\pi^2M_{\rm
SQ}^2I_B^2\tan^2\gamma_{\rm in,0}\frac{Z_{\rm eff}}{i\omega\Phi_0^2}I_{\rm Q}
\end{equation}
where $\delta\Phi_{\rm cl}\simeq M_{\rm SQ}  I_{c,0}\cos\gamma_{\rm ex,0}\sin
\gamma_{\rm in,0}$ is the non-fluctuating back-action from the
SQUID.

From the two-dimensional problem, we can now restrict ourselves to the
one-dimensional subspace defined by the preferred tunneling direction
\cite{Hans}, which is described by an effective phase $\varphi$.
The potential restricted on this direction, $U_{\rm 1D}(\varphi)$ 
has the form of a  double
well \cite{LeggettReview,Weiss} with stable minima situated at
$\pm\varphi_0$. In  this way, we can expand
$U_{\rm 1D}(\varphi,\Phi_{\rm q})\simeq U(\varphi,\Phi_q,x)+I_{\rm
Q}(\varphi)I_{\rm Q}M_{\rm SQ}$.  
Approximating the
phase-dependence of the circulating current as
$I_{\rm Q}(\varphi)\approx I_{\rm p}\varphi/\varphi_0$ where $I_{\rm p}$ 
the circulating current
in one of the stable minima of $\varphi$, we end 
up with the classical equation of motion of the qubit including the
backaction and the friction induced from the SQUID 
\begin{eqnarray}
& & \left[-C_{\rm
eff}\left(\frac{\Phi_0}{2\pi}\right)^2\omega^2+2\pi^2M_{\rm
SQ}^2I_{\rm bias}^2\tan^2\gamma_{\rm in,0}\frac{Z_{\rm eff}I^2_{\rm
p}}{i\varphi_0\omega\Phi_0^2}\right] \varphi \nonumber \\
& & = -\partial_\varphi
U_{\rm 1D}(\varphi,\Phi_{x,q}+\delta\Phi_{\rm cl}).
\end{eqnarray}
From this form, encoded as $D(\omega)\varphi(\omega)=-\partial
U/\partial\varphi$ we can use the prescription given  in \cite{Leggett} and
identify the spectral function for the continuous, classical model as
$J_{\rm cont}={\rm Im} D(\omega)$. From there, we can do the two-state
approximation for the particle in a double well \cite{Weiss} and find
$J(\omega)$ in analogy to \cite{EPJB}
\begin{equation}
J(\omega )=\frac{\left( 2\pi \right) ^{2}}{\hbar\omega
}\left( \frac{M_{\rm SQ}I_{\rm p}}{\Phi _{0}}\right) ^{2}I_{\rm bias}^{2}\tan
^{2} \left(\frac{\pi\Phi}{\Phi_0}\right)\; {\rm Re} \lbrace Z_{\rm eff}(\omega )\rbrace .
\label{Jwsquid}
\end{equation}

\subsection{Qubit Dynamics under the Influence of Decoherence}

From $J(\omega)$, we can analyze the dynamics of the system by studying
the
reduced density
matrix, i.e.\ the density matrix of the full system where the details of
the environment have been integrated out, by a number of different methods. The low damping limit,  
$J(\omega)/\omega\ll 1$ for all frequencies, is most desirable for
quantum computation.  Thus, the energy-eigenstates of the qubit
Hamiltonian, Eq.\ (\ref{2lsHam}), are  the appropriate starting point
of our discussion.  In this case, the relaxation rate $\Gamma _{r}$
(and relaxation time $\tau _{r}$) are determined by the environmental
spectral function $J(\omega )$ at the frequency of the level separation
$\nu $ of the qubit
\begin{equation}
\Gamma _{r}=\tau _{r}^{-1}=\frac{1}{2}\left( \frac{\Delta }{\nu
}\right) ^{2}J\left(\frac{\nu}{\hbar}\right)\coth \left( \frac{\nu }{2k_{B}T}\right) ,
\label{Gmix}
\end{equation}
where $T$ is the temperature of the bath. The dephasing rate $\Gamma
_{\phi } $ (and dephasing time $\tau _{\phi }$) is
\begin{equation}
\Gamma _{\phi }=\tau _{\phi }^{-1}=\frac{\Gamma _{r}}{2}+
2\pi\alpha\left( \frac{\varepsilon }{\nu }\right) ^{2}\,
\frac{k_{B}T}{\hbar }
\label{Gphase}
\end{equation}
with $\alpha=\lim_{\omega\rightarrow0} J(\omega)/(2\pi\omega)$. 
These expressions have been derived in the context of NMR
\cite{abragam61} and recently been
confirmed by a full  path-integral analysis \cite{grifoni99}. In this paper,
all rates are calculated for this regime.

For performing efficient measurement, one can afford to go to the
strong damping regime. A well-known  approach to this problem, the
noninteracting blip approximation (NIBA) has been derived in Ref.\
\cite{Leggett}. This approximation gives good predictions at
degeneracy, $\epsilon=0$. At low $|\epsilon|>0$ it contains an artifact
predicting incoherent dynamics even at
weak damping. At high bias, $\epsilon\gg\Delta$ and at strong damping,
it becomes asymptotically correct again. We will not detail this
approach here more, as it has been extensively covered in
\cite{LeggettReview,Weiss}.

If $J(\omega)$ is not smooth but contains strong
peaks the situation becomes more involved: At some frequencies,
$J(\omega)$ may fall in the weak and at others in the strong
damping limit. In some cases, whern
 $J(\omega)\ll \omega$ holds at least for
$\omega\le\Omega$ with some $\Omega\gg\nu/\hbar$, this can be treated
approximately:
one can first 
renormalize $\Delta_{\rm eff}$ through the high-frequency
contributions \cite{LeggettReview} and then perform a weak-damping
approximation from the fixed-point Hamiltonian. This is
detailed in Ref.\ \cite{Neumann}. In the general case, more
involved methods such as flow equation
renormalization \cite{Kleff} have to be used.

\section{Engineering the Measurement Apparatus}

From Eq. \ref{Jwsquid} we see that engineering the decoherence
induced by the measurement apparatus essentially means engineering
$Z_{\rm eff}$. This includes also the
contributions due to the measurement apparatus. 
In this section, we are going to outline and compare
several options suggested in literature.  We assume a perfect current
source that ramps the bias current $I_{\rm bias}$ through the SQUID. The fact
that the current source is non-ideal, and that the wiring to the SQUID
chip has an impedance is all modeled by the impedance $Z(\omega)$. The
wiring can be engineered such that for a very wide frequency range the
impedance $Z(\omega)$ is on the order of the vacuum impedance, and can be
modeled by its real part $R_{l}$. It typically has a value of
$100\;{\rm \Omega} $.

\subsection{R-Shunt}

It has been suggested
\cite{Makhlin2}  to overdamp the SQUID by 
making the shunt circuit a simple resistor
$Z(\omega)=R_{\rm S}$ with 
$R_{\rm S}\ll \sqrt{L_{\rm kin}/2C_{\rm J}}$. This is inspired by an analogous
setup for charge qubits, \cite{Makhlin}. 
Following the parameters given in \cite{EPJB}, a SQUID with $I_{\rm c,0}=
200$nA at $\Phi/\Phi_0\simeq 0.75$ biased at $I_{\rm bias}=120$nA, 
we find
$L_{\rm kin}\simeq 2\cdot 10^{-9} H$. Together with $C_{\rm J}\simeq$
1fF, this means that the SQUID is overdamped if $R\ll R_{\rm max}=1.4
{\rm k\Omega}$. 
Using Eq.\
\ref{Jwsquid}, we find that this provides an Ohmic environment with
Drude-cutoff, $J(\omega)=\alpha\omega/(1+\omega^2/\omega_{LR}^2)$
where $\omega_{LR}=R/L_{\rm kin}$ and $\alpha=(2\pi)^2/\hbar
\left(M_{SQ}I_q/\Phi_0\right)^2I_{\rm bias}^2\tan^2
(\pi\Phi/\Phi_0)L^2_{\rm kin}/R_{\rm S}$. Using the parameters
from Ref. \cite{EPJB},  $M_{\rm SQ}I_q/\Phi_0=0.002$, we find $\alpha
R = 0.08{\rm \Omega}$ and $\omega_{\rm LR}/R=8.3 {\rm GHz/\Omega}$. Thus, for
our range of parameters (which essentially correspond to weak coupling
between SQUID and qubit), one still has {\em low} damping of the qubit from the 
(internally
overdamped) environment
at reasonable shunt resistances down to tens of Ohms. For
such a setup, one can apply the continuous weak measurement theory as
it is outlined e.g.\ in \cite{Makhlin2}. This way, one can readily 
describe the readout through measurement of $Z_{\rm eff}$ which leaves
the system on the superconducting branch.
If one desires to read out
the state by monitoring the voltage at bias currents {\em above} the
$I_{\rm c, eff}$, our analysis only describes the pre-measurement phase and
at least shows that the system is
hardly disturbed when the current is ramped.
\begin{figure}[htb]
\begin{minipage}{0.99\columnwidth}
%h=here, t=top, b=bottom, p=separate figure page
%\begin{center}\leavevmode
\vspace{5mm}
\includegraphics[width=0.8\columnwidth]{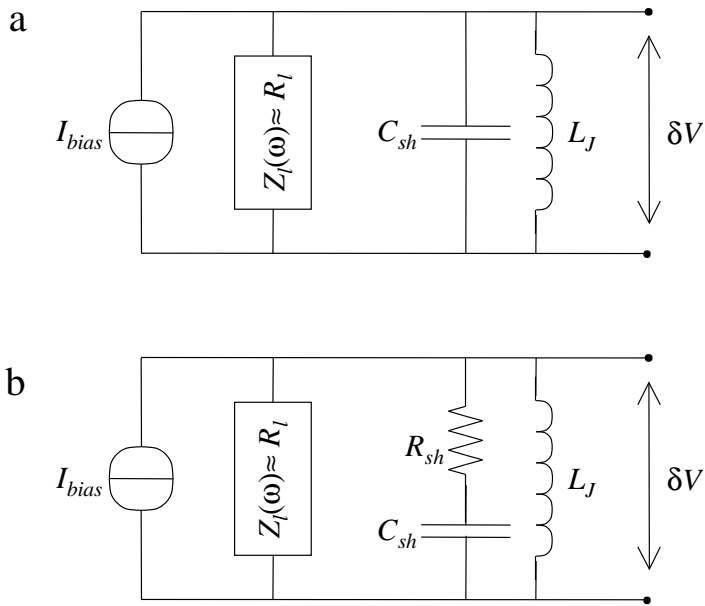}
\caption{Circuit models for the $C$-shunted DC-SQUID (a) and the
$RC$-shunted DC-SQUID (b). The SQUID is modeled as an inductance
$L_J$. A shunt circuit, the superconducting capacitor $C_{\rm sh}$ or
the $R_{\rm sh}$-$C_{\rm sh}$ series, is fabricated on chip very close to
the SQUID. The noise that couples to the qubit results from
Johnson-Nyquist voltage noise $\protect \delta V$ from the
circuit's total impedance $Z_{\rm eff}$. $Z_{\rm eff}$ is formed by a parallel
combination of the impedances of the leads $Z_l$, the shunt and
the SQUID, such that $Z_{\rm eff}^{-1}=1/Z_l+1/(R_{sh}+1/i \protect\omega
C_{sh}) + 1/i \protect\omega L_J$, with $R_{sh} = 0$
for the circuit (a)} \label{figsqcir}
%\end{center}
\end{minipage}
\end{figure}

\subsection{Capacitive Shunt}
 
\begin{figure}[htb]
%h=here, t=top, b=bottom, p=separate figure page
%\begin{center}\leavevmode
\begin{minipage}{0.99\columnwidth}
\includegraphics[width=0.8\columnwidth]{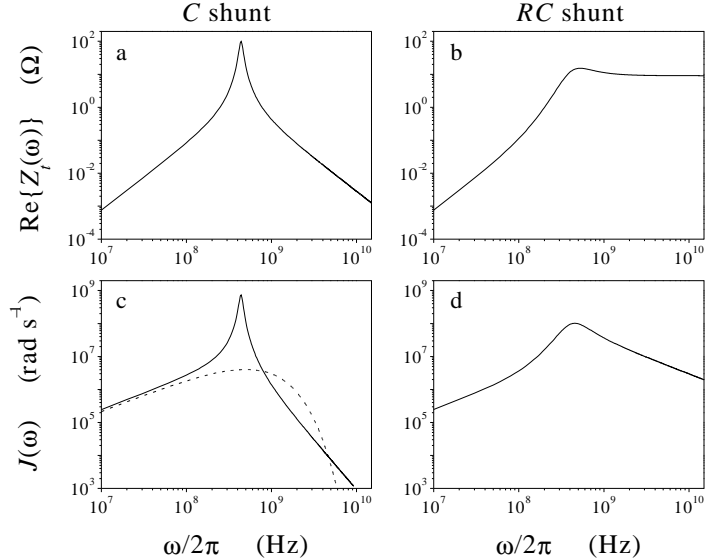}
\caption{A typical ${\rm Re}\{Z_{t}(\protect\omega)\}$ for the
$C$-shunted SQUID (a) and the $RC$-shunted SQUID (b), and
corresponding $J(\protect \omega)$ in (c) and (d) respectively.
For comparison, the dashed line in (c) shows a simple Ohmic
spectrum, $J(\omega)=\alpha\omega$ with exponential cut off $\protect\omega_c
/ 2\protect\pi$ = 0.5 GHz and $\protect\alpha$ = 0.00062. The
parameters used here are $I_{\rm p} = 500$ nA and $T = 30$ mK. The
SQUID with $2I_{co}$ = 200 nA is operated at $f = 0.75 \,
\protect\pi $ and current biased at 120 nA, a typical value for
switching of the $C$-shunted circuit (the $RC$-shunted circuit
switches at higher current values). The mutual inductance $M$ = 8
pH (i.~e.~$MI_{\rm p}/\Phi_{0} = 0.002$). The shunt is
$C_{\rm sh}$ = 30 pF and for the $RC$ shunt $R_{\rm sh} = $ 10 ${\rm \Omega}$.
The leads are modeled by $R_{l} = 100$ ${\rm \Omega} $}
\label{figsqrez}
%\end{center}
\end{minipage}
\end{figure}
Next, we consider a large superconducting capacitive shunt
(Fig.~\ref{figsqcir}a, as implemented in Refs.\ \cite{Irinel,Caspar}).
The $C$ shunt only makes the effective mass of the SQUID's external
phase $\gamma_{\rm ex}$ very heavy. 
The total impedance $Z_{\rm eff}(\omega)$ and $J(\omega)$ 
are modeled as before, 
see Fig.\ \ref{figsqrez}.
As limiting values, we find
\begin{equation}
{\rm Re} \{Z_{\rm eff}(\omega)\}\approx
 \left \{
\begin{array}{cc}
\frac{\omega ^{2}L_{J}^{2}}{R_{l}}, & {\rm for}\;\omega \ll \omega
_{LC} \\ R_{l}, & {\rm for}\;\omega =\omega _{LC} \\ \frac{1}{\omega
^{2}C_{sh}^{2}R_{l}}, & {\rm for}\;\omega \gg \omega _{LC}
\end{array}
 \right.
\end{equation}
We can
observe that this circuit is a weakly damped $LC$-oscillator and
it is clear from (\ref{Gmix}) and (\ref{Jwsquid})
that one should keep its resonance frequency 
$\omega_{\rm LC}=1/\sqrt{L_{\rm J}C_{\rm sh}}$, where ${\rm Re} \{Z_{\rm eff}(\omega )\}$ has a
maximum, away from the qubit's resonance $\omega _{\rm res}=\nu /\hbar
$. This is usually done by chosing 
$\omega _{\rm LC}\ll \omega_{\rm res}$. For a $C$-shunted circuit with $\omega _{LC}\ll \omega_{res}$, 
this yields for $J(\omega\approx \omega_{\rm LC})$ 
\begin{equation}
J(\omega)\approx \frac{\left( 2\pi \right) ^{2}}{\hbar\omega^{3}}\left( \frac{MI_{\rm p}}{\Phi _{0}}\right)
^{2}I_{\rm bias}^{2}\tan ^{2}\left(\frac{\pi\Phi}{\Phi_0}\right)\;\frac{1}{C_{sh}^{2}R_{l}}
\end{equation}
The factor $1/\omega ^{3}$ indicates a natural cut-off for $J(\omega)
$, which prevents the ultraviolet divergence
\cite{LeggettReview,grifoni99} and which in much of the theoretical
literature is introduced by hand. Using Eq. \ref{Gmix}, we can
directly analyze mixing
times $\tau _{r}$ vs $\omega _{\rm res}$ for typical sample parameters
(here calculated with the non-approximated version of ${\rm Re}
\{Z_{t}(\omega)\}$), see Ref.\ \cite{EPJB} for details. The mixing rate 
is then
$\Gamma _{r}\approx \left( 2\pi \Delta /\hbar \right) ^{2} \omega
_{res}^{-5} \left(
MI_{p}/\Phi _{0}\right) ^{2} I_{\rm bias}^{2}\tan ^{2} (\pi\Phi/\Phi_0)
( 2 \hbar C_{sh}^{2}R_{l} )^{-1} \coth \left( \hbar \omega
_{res}/2k_{B}T \right)$ .
With the $C$-shunted circuit it seems possible to
get $\tau _{r}$ values that are very long. They are compatible with
the ramp times of the SQUID, but too slow for fast repetition rates.
  For the parameters used here they
are in the range of 15 ${\rm \mu s}$.  While this value is close to
the desired order of magnitude,  one has to be aware of the fact that at
these high switching current values the linearization of the junction
as a kinetic inductor
may underestimate the actual noise.  In that regime, phase diffusion
between different minima of the  washboard potential also becomes
relevant and changes the noise properties \cite{joyez99,coffey96}.

\subsection{RC-Shunt}

As an alternative we will consider a shunt that is a series
combination of a capacitor and a resistor
(Fig.~\ref{figsqcir}b) ($RC$-shunted
SQUID).  The $RC$ shunt also adds damping at the plasma frequency of
the SQUID, which is needed for realizing a high resolution of the
SQUID readout (i.~e.~for narrow switching-current histograms)
\cite{joyez99}.  The total impedance $Z_{t}(\omega)$ of the two
measurement circuits are modeled as in Fig.~\ref{figsqcir}.  For the
circuit with the $RC$ shunt
\begin{equation}
{\rm Re} \{Z_{t}(\omega)\}\approx \left \{
\begin{array}{cc}
\frac{\omega ^{2}L_{J}^{2}}{R_{l}}, & {\rm for}\;\omega \ll \omega
_{LC}\;\;\;\;\;\;\;\;\;\; \\ \le R_{l}, & {\rm for}\;\omega =\omega
_{LC}\ll \frac{1}{R_{sh}C_{sh}} \\ R_{l}//R_{sh}, & {\rm for}\;\omega
=\omega _{LC}\gg \frac{1}{R_{sh}C_{sh}} \\ R_{l}//R_{sh}, & {\rm
for}\;\omega \gg \omega _{LC}\;\;\;\;\;\;\;\;\;\;
\end{array}
 \right.
\end{equation}
The difference mainly concerns frequencies $\omega >\omega _{LC}$,
where the $C$-shunted circuit has a stronger cutoff in 
${\rm Re} \{Z_{\rm eff}(\omega )\},$ and
thereby a relaxation rate, that is several orders lower than for the
$RC$-shunted circuit. Given the values of $J(\omega)$ from 
Fig.~\ref{figsqrez} one can directly see from the values of
that an $RC$-shunted circuit with otherwise similar
parameters yields at $\omega _{res}/2\pi =$ 10 GHz relaxation times
that are about four orders of magnitude shorter.

\section{Coupled Qubits}

So far, we have applied our modeling 
only to single qubits. In order to study entanglement in
a controlled way and to eventually perform quantum algorithms, this
has to be extended to coupled qubits.

\subsection{Hamiltonian}

There is a number of ways how to couple two solid-state qubits in a
way which permits universal quantum compuation. If the qubit states
are given through real spins, one typically obtains a
Heisenberg-type exchange  coupling. For other qubits, the
three components of the pseudo-spin typically correspond to physically
completely distinct variables. In our case, $\hat{\sigma}_z$
corresponds to the flux through the loop whereas $\hat{\sigma}_{\rm
x/y}$ are charges. Consequently, one usually finds 
Ising-type couplings. The case of $\hat{\sigma}_{\rm
y}^{(1)}\otimes\hat{\sigma}^{(2)}_{\rm y}$ coupling, 
i.e.\ coupling by a component which
is orthogonal to all possible single-qubit Hamiltonians, has been
extensively studied \cite{Michele,thorwart}, because this type is
straightforwardly realized as a {\em tunable} coupling of charge qubits
\cite{Makhlin}. We study the generic case of coupling  the
``natural'' variables of the pseudospin to each other, which can be
realized in flux qubits using a switchable superconducting transformer
\cite{Hans,hannes}, but has also been experimentally utilized for
coupling charge qubits by fixed capacitive interaction
\cite{Nakamura}.

We model the Hamiltonian of a system of two qubits, coupled via
Ising-type coupling.  Each of the two qubits is described 
by the Hamiltonian Eq.\ (\ref{2lsHam}). 
The coupling between the qubits is described by $\hat{H}_{\rm qq}=-(K/2)
\hat{\sigma}_z^{(1)}\otimes  \hat{\sigma}_z^{(2)}$ that represents
e.g.\ inductive interaction. Thus, the complete two-qubit
Hamiltonian in the absence of a dissipative environment reads
\begin{equation} \label{Hop2qb_nobath}
\hat{H}_{2qb} = - \frac{1}{2} \sum_{i=1,2}\left( \epsilon_i \hat
\sigma_z^{(i)} + \Delta_i \hat \sigma_x^{(i)} \right) -
\frac{1}{2} K \hat \sigma_z^{(1)} \hat \sigma_z^{(2)}\textrm{.}
\end{equation}
For two qubits, there are several ways to couple to the environment:
Both qubits may couple to a common bath such as 
picked up by coupling elements \cite{Hans}. 
Local
readout and control electronics coupling to individual qubits
\cite{Hans} can be described as coupling to two uncorrelated baths. In 
analogy to the procedure described above, one can determine the spectral
functions of these baths by investigating the corresponding impedances. 

In the case of two uncorrelated baths, the full Hamiltonian reads
\begin{equation} \label{Hop_2baths}
\hat{H}_{2qb}^{2b} =\hat{H}_{2qb}+
\sum_{i=1,2} \frac{1}{2} \hat \sigma_z^{(i)} \widehat{X}^{(i)} +
\hat{H}_{B_1} + \hat{H}_{B_2}\textrm{,}
\end{equation}
$\widehat{X}^{(i)}=\zeta
\sum_\nu \lambda_\nu x_\nu$ are collective coordinates of the bath.  
In the
case of two qubits coupling to one common bath we model our two qubit
system in a similar way with the Hamiltonian
\begin{equation} \label{Hop_1bath}
\hat{H}_{2qb}^{1b}  = \hat{H}_{2qb}
+ \frac{1}{2} \left( \hat \sigma_z^{(1)} + \hat \sigma_z^{(2)}
\right) \widehat{X}  + \hat{H}_{B}
\end{equation} 
where $\hat{X}$ is a collective bath coordinate similar to above. 

\subsection{Rates}

\begin{figure}[ht]
\begin{center}
\includegraphics*[width=0.8\columnwidth]{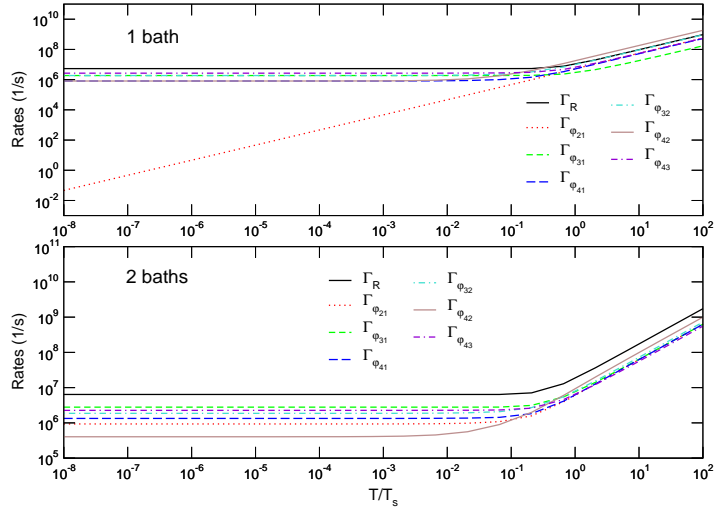}
\end{center}
\caption[Log-log plot of the temperature dependence of the sum of the
four relaxation rates and selected dephasing rates.]{Log-log plot of
the temperature dependence of the sum of the four relaxation rates and
selected dephasing rates. Qubit parameters $K$, $\epsilon$ and $\eta$
are all set to $E_s$ and the bath is assumed to be Ohmic 
$\alpha=10^{-3}$. The upper panel shows the
case of one common bath, the lower panel the case of two distinct
baths. At the characteristic temperature of approximately $0.1\cdot
T_s$ the rates increase very steeply}
\label{fig_temp_dependence_1}
\end{figure}

We can derive formulae
for relaxation and dephasing rates similar to Eqs.\
(\ref{Gmix})  and (\ref{Gphase}).  Our Hilbert space is now four-dimensional. 
We label the
eigenstates as $\ket{E1} \dots \ket{E4}$. We chose
$\ket{E1}$ to be the singlet state $\left(\ket{\up\down}-\ket{\down\up}\right)/\sqrt{2}$, which is always an eigenstate \cite{MarkusFrankPRA} whereas $\ket{E2}\dots\ket{E4}$ are the energy eigenstates in 
the triplet subspace, which are typically {\em not} the eigenstates of
$\hat{\sigma}_z^{(1)}+\hat{\sigma}_z^{(2)}$. As we have 4 levels, 
we have 6 independent possible quantum coherent oscillations,
each of which has its own dephasing rate, as well as 4
relaxation channels, one of which has a vanishing rate indicating the
existence of a stable thermal equilibrium point. The expressions for
the rates, although of similar form as in Eqs.\ (\ref{Gmix}) and (\ref{Gphase}) 
are
rather involved and are shown in \cite{MarkusFrankPRA}.  Figure
\ref{fig_temp_dependence_1} displays the dependence of typical
dephasing rates and the sum of all relaxation rates $\Gamma_R$ on 
temperature for the case $\Delta=\epsilon=K=h\nu_S$ with $\nu_S=1{\rm GHz}$. 
The rates
are of the same magnitude for the case of one common bath and two
distinct baths.  If the temperature is increased above the roll off point
set by the intrinsic energy scales, $T_s = (h/k_B) \nu_s =
4.8 \cdot 10^{-2}\textrm{ K}$, where $E_s = 1$GHz, the increase of the dephasing and
relaxation rates follows a linear dependence, indicating that the
environmental fluctuations are predominantly thermal.
 As a notable exception, in the case of one common
bath the dephasing rates $\Gamma_{\varphi_{21}}=\Gamma_{\varphi_{12}}$
go to zero when the temperature  is decreased while all other rates
saturate for $T \rightarrow 0$.
This can be
understood as follows: the singlet state $\ket{E1}$
is left invariant by the Hamiltonian of coupled
qubits in a common bath, Eq.\ (\ref{Hop_1bath}), i.e.\ it is 
an energy eigenstate left
unaffected by the environment. Superpositions
of the singlet with {\em another} eigenstate are usually
still unstable, because the other eigenstate generally suffers from 
decoherence. However, the lowest-energy state 
of the triplet subspace $\ket{E2}$ cannot decay by 
spontaneous emission and flip-less dephasing vanishes at $T=0$, 
hence the dephasing rate between eigenstates $\ket{E1}$ and $\ket{E2}$ vanishes 
at low temperatures, see Fig.\ 
\ref{fig_temp_dependence_1}. As shown in \cite{MarkusFrankPRA}, 
there can be more
``protected'' transitions of this kind if the qubit parameters are adjusted
such that the symmetry between the unperturbed qubit and the coupling to
the bath is even higher, e.g.\ at the working point for a CPHASE operation.

\subsection{Gate Performance}
The rates derived in the previous section are numerous and do strongly 
dependend on the tunable parameters of the qubit. 
Thus, they do not yet allow a full
assesment of the performance as a quantum logic element. 
A quantitative measure of how well a two-qubit 
setup performs a quantum logic gate operation are the
gate quality factors introduced in \cite{GateQualityFactors}:
the fidelity, purity, quantum degree and entanglement capability. These
factors characterize the density matrices obtained after attempting
to perform the gate operation in a hostile environment, starting from
all possible initial conditions
$\rho(0)=\ket{\Psi_{in}^j}\bra{\Psi_{in}^j}$. To form all possible initial density matrices needed to calculate the
gate quality factors, we use the 16 unentangled product states
$\ket{\Psi_{in}^j}$, $j=1,\dots ,16$ defined \cite{thorwart} according
to $\ket{\Psi_a}_1\ket{\Psi_b}_2$, ($a,b=1,\dots,4$),  with
$\ket{\Psi_1}=\ket{0}$, $\ket{\Psi_2}=\ket{1}$,
$\ket{\Psi_3}=(1/\sqrt{2})(\ket{0}+ \ket{1})$, and
$\ket{\Psi_4}=(1/\sqrt{2}) (\ket{0}+i\ket{1})$. They form one possible
basis set for the superoperator $\nu_{G}$ which describes the
open system dynamics such that  $\rho(t_{G})=\nu_{G}
\rho(0)$    \cite{thorwart,GateQualityFactors}. The CNOT gate is
implemented using rectangular DC pulses and describing
dissipation through the Bloch-Redfield equation as described in
\cite{Makhlin,MarkusFrankPRA}. 

The fidelity is defined as
$\mathcal{F} = (1/16) \sum_{j=1}^{16}
\braket{\Psi_{in}^j|U_{G}^+\rho^j_{G}U_{G}| \Psi^j_{in}}\textrm{.}$
The fidelity is a
measure of how well a quantum logic operation was performed. 
Clearly, the fidelity for the ideal quantum
gate operation is equal to 1.
The second quantifier is the purity
$\mathcal{P} = (1/16) \sum_{j=1}^{16}
\mbox{tr}\left[(\rho^j_{G})^2\right]\textrm{,}$
which should be 1 in a pure and $1/4$ in a fully mixed
state.  The purity characterizes the effects of decoherence.
The quantum degree measures nonlocality. It 
is defined as the maximum
overlap of the resulting density matrix after the quantum gate
operation with the maximally entangled Bell-states
$\mathcal{Q} = \max_{j,k} \braket{\Psi_{me}^k|\rho_{G}^j|\Psi_{me}^k}$.
For an ideal entangling operation, e.g. the CNOT gate, the quantum
degree should be 1. It
has been shown \cite{nonlocality} that all density operators that have
an overlap with a maximally entangled state that is larger than the
value $0.78$ \cite{thorwart} violate the Clauser-Horne-Shimony-Holt
(CHSH) inequality and are thus non-local.
The  entanglement capability $\mathcal{C}$ is
the smallest eigenvalue of the partially transposed density matrix for
all possible unentangled input states $\ket{\Psi_{in}^j}$. (see
below).  It has been shown \cite{peres} to be negative for an
entangled state. This quantifier should be -0.5, e.g. for the ideal
CNOT, thus characterizing a maximally entangled final state. 

\begin{figure}[t]
\begin{center}
\includegraphics*[width=0.8\columnwidth]{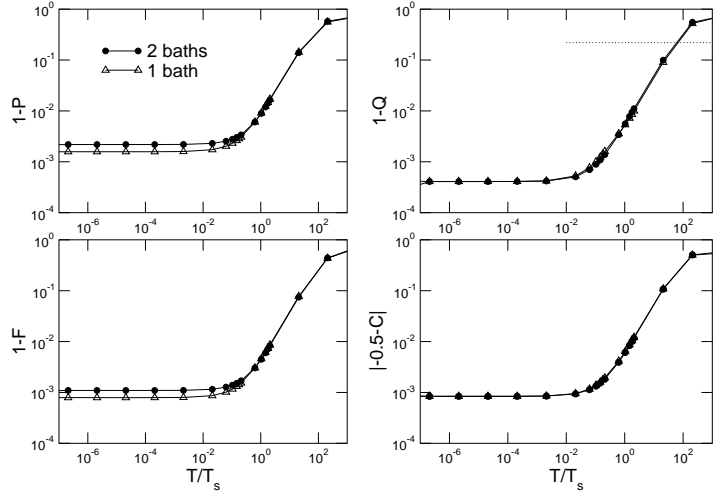}
\end{center}
\caption[Log-log plot of the temperature dependence of the deviations
of the four gate quantifiers from their ideal values.]{Log-log plot of
the temperature dependence of the deviations of the four gate
quantifiers from their ideal values. Here the temperature is varied
from $\approx 0 $  to  $2\cdot E_s$. In all cases
$\alpha=\alpha_1=\alpha_2=10^{-3}$. The dotted line indicates the
upper bound set by the Clauser-Horne-Shimony-Holt inequality}
\label{gqfTdep}
\end{figure}

In Fig.\ \ref{gqfTdep}, the deviations due to decoherence of the 
gate quality factors
from their ideal values are shown. Similar to most of the
rates, all gate quality factors saturate at temperatures below a 
threshold set by the qubit energy scales. The deviations
grow linearily at higher temperatures until they reach
their theoretical maximum.
Comparing the different coupling scenarios, we see that
at low temperatures, the purity
and fidelity are higher for the case of one common bath, but if
temperature is increased above this threshold, fidelity
and purity are approximately equal for both the case of one common and
two distinct baths.
This is related to the fact that in the case of one common bath all relaxation
and dephasing rates vanish during the two-qubit-step of the CNOT (see
\cite{MarkusFrankPRA} for details),  due
to the special symmetries of the Hamiltonian, when temperature goes to
zero as discussed above. Still, the quantum degree
and the entanglement capability tend towards the same value for both
the case of one common and two distinct baths. This is due to the fact
that both quantum degree and entanglement capability are, different
than fidelity and purity, not defined as mean values but rather
characterize the ``best'' possible case of all given input
states. 

In the recent work by Thorwart and H\"anggi \cite{thorwart}, the CNOT
gate was investigated for a $\hat{\sigma}_y^{(i)}\otimes\hat{\sigma}_y^{(j)}$
coupling scheme  and one common bath. They find a pronounced
degradation of the gate performance with gate quality
factors only weakly depending on temperature. If we set the dissipation
and the intrinsic energy scale to the same values as in their work, 
we also observe only a weak decrease of the gate
quality factors for both the case of one common bath and two distinct
baths in the same temperature range discussed by Thorwart and
H\"anggi. However, see Fig.\ \ref{gqfTdep}, overall we find 
substantially better values.  This is due to the fact that for
$\hat{\sigma}_y\otimes\hat{\sigma}_y$ coupling, the Hamiltonian does {\em not}
commute with the coupling to the bath during the two-qubit steps of
the pulse sequence, i.e.\ the symmetries of the coupling to the bath
and the inter-qubit coupling are not compatible.
The dotted line in Fig.~\ref{gqfTdep} shows that already at comparedly high 
temperature, about 20 qubit energies, a
quantum degree larger than $\mathcal{Q} \approx
0.78$ can be achieved. Only then, the Clauser-Horne-Shimony-Holt inequality is
violated and non-local correlations between the qubits occur as
described in \cite{thorwart}. Thus, even under rather modest requirements
on the experimental setup which seem to be feasible with present
day technology, it appears to be possible to demonstrate nonlocality
and entanglement between superconducting flux qubits. 

\section{Summary}

It has been outlined, how one can model the decoherence of an
electromagnetic environment inductively coupled to a superconducting flux
qubit. We have exemplified a procedure based on analyzing the
classical friction induced by the  environment for the specific case
of the read-out SQUID. It is shown that the SQUID can be effectively
decoupled from the qubit if no bias current is applied. The
effect of the decoherence on relaxation and dephasing rates
of single qubits has been discussed 
as well as the gate performance of coupled qubits. We
have shown that by carefully engineering the impedance and the
symmetry of the coupling, one can reach excellent gate quality which
complies with the demands of quantum computation.

We would like to thank M.\ Governale, T.\ Robinson,  and M.\ Thorwart
 for discussions.  FKW and MJS acknowledge support from ARO under
 contract-No. P-43385-PH-QC.

%INDEX%%%%%%%%%%%%%%%%%%%%%%%%%%%%%%%%%%%%%%%%%%%%%%%%%%%%%%%%%%%%%%%
% Please code your entries to include a "mutual" subject index in the
% standard syntax. For your own purposes you may print your
% "personal" index by using the following commands:
%
%\clearpage
%\addcontentsline{toc}{section}{Index}
%\flushbottom
%\printindex
%%%%%%%%%%%%%%%%%%%%%%%%%%%%%%%%%%%%%%%%%%%%%%%%%%%%%%%%%%%%%%%%%%%%%

\end{document}